\documentclass[twocolumn]{aastex631}

\usepackage{amsmath}
\usepackage{amssymb}
\usepackage{bm}
\usepackage{booktabs}
\usepackage{graphicx}
\usepackage{xcolor}

\newcommand{\Msun}{M_\odot}

\newcommand{\phat}{\hat p}
\newcommand{\ehat}{\hat\epsilon}
\newcommand{\wideinlinefigure}[4]{%
  \begin{figure*}[t]
  \centering
  \includegraphics[width=#2]{#3}
  \caption{#4}
  \label{#1}
  \end{figure*}
}

\begin{document}
\title{An Analytical Toy Equation of State for Neutron Stars Consistent with Current Observations}
\shorttitle{Analytical Toy EOS}
\shortauthors{Chen, Zhou, and Zhang}

\author[0009-0004-4744-4597]{Tian-Shun Chen}
\email{Delta\_Chen@163.com}
\affiliation{Tsung-Dao Lee Institute, Shanghai Jiao-Tong University, Shanghai 201210, China}
\affiliation{Department of Physics and Institute for Quantum Science and Technology, Shanghai University, Shanghai 200444, China}

\author[0009-0002-8381-7130]{Xiao-Ding Zhou}
\email{18038288465@shu.edu.cn}
\affiliation{Department of Physics and Institute for Quantum Science and Technology, Shanghai University, Shanghai 200444, China}

\author[0000-0001-6776-9074]{Kilar Zhang\textsuperscript{*}}
\email{kilar@shu.edu.cn}
\altaffiliation{Corresponding author}
\affiliation{Department of Physics and Institute for Quantum Science and Technology, Shanghai University, Shanghai 200444, China}
\affiliation{Shanghai Key Laboratory of High Temperature Superconductors, Shanghai 200444, China}

\begin{abstract}
Fast analytic and semi-analytic studies of neutron stars often require an equation of state that is convenient to evaluate while producing relativistic stellar sequences compatible with current multimessenger constraints.  We construct such a benchmark by scanning a smooth double polytropic relation for the energy density as a function of pressure, \(\ehat(\phat)=a_1\phat^{\Gamma_1}+a_2\phat^{\Gamma_2}\).  The parameters are selected with filters based on massive pulsars, tidal deformability from the binary-neutron-star event GW170817, and NICER mass-radius measurements.  A single polytropic baseline scan finds no model passing all filters, whereas a double-polytrope scan identifies a viable region.  A curve-integral score, evaluated against public NICER and GW170817 posterior data sets, is then used to choose benchmark equations of state within this region.  The selected representatives support \(M_{\max}=2.44\)--\(2.49\,M_\odot\), with \(R_{1.4}\simeq11.3\) km and \(\Lambda_{1.4}=485\)--512, and remain causal on the stable branch.  This compact analytic family provides reference cases for relativistic stellar-structure tests at current observational scales.
\end{abstract}

\section{Introduction}
\label{sec:introduction}

The combination of two-solar-mass pulsars, binary-neutron-star
gravitational-wave (GW) observations, and pulse-profile modeling has raised the
standard for neutron-star test equations of state: a useful benchmark should
remain simple while matching current mass, radius, and tidal scales.  Radio
timing gives robust high-mass
anchors, including
PSR~J0740+6620 with \(M=2.08\pm0.07\,\Msun\) \cite{Fonseca2021}; earlier
systems such as PSR~J1614--2230 and PSR~J0348+0432 established the
approximately two-solar-mass scale \cite{Demorest2010,Antoniadis2013}.  The
binary-neutron-star event GW170817 supplied tidal-deformability information
\cite{Abbott2017GW170817,Abbott2018EOS}, and Neutron Star Interior Composition
Explorer (NICER) pulse-profile studies now provide mass-radius constraints for
both canonical and high-mass pulsars
\cite{Riley2019,Miller2019,Riley2021,Miller2021,Dittmann2024,Choudhury2024}.

We therefore develop an analytic benchmark for calculations in which repeated
equation-of-state (EOS) evaluations and transparent numerical checks are
important.  The model is required to be low dimensional, differentiable,
physically self-consistent, and able to provide the usual relativistic
diagnostics: mass-radius sequences, tidal deformabilities, and sound speeds.

Low-dimensional EOS representations are well established in neutron-star
phenomenology and inference.  Reviews of dense-matter constraints emphasize the
complementarity of nuclear physics, high-mass pulsars, radii, and tidal
measurements \cite{OzelFreire2016,Lattimer2012}.  Piecewise polytropes provide
a compact parametrization of cold dense matter \cite{Read2009}; spectral and
causal spectral representations improve smoothness and physical control
\cite{Lindblom2010,Lindblom2018}; generalized piecewise polytropes can enforce
continuous sound speed \cite{OBoyle2020}; analytical representations of
unified EOS tables make microphysical models easier to deploy
\cite{Potekhin2013}; and modern fits to unified tabulated EOSs retain
polytropic efficiency while tracking crust-core consistency
\cite{Suleiman2022}.  Continuous-sound-speed parametrizations have also been
developed for neutron-star simulations \cite{Servignat2024}, while Bayesian
and phenomenological studies combine radio, X-ray, and gravitational-wave
information across broader EOS families
\cite{Steiner2010,Hebeler2013,LandryEssick2019,Annala2020,Chimanski2023}.
The dependence of EOS conclusions on parameterization and prior choice has
been emphasized in studies of piecewise-polytropic and phenomenological
inference \cite{Raaijmakers2018,Greif2019,Legred2022}.
Large phenomenological EOS ensembles further support updated tidal universal
relations \cite{Godzieba2021}, and recent effective-field-theory (EFT)
calculations of dynamical tidal response have used single- and two-term
polytropic fits to tabulated baryonic EOSs \cite{Jarequi2026}. 

In this setting, we construct a compact analytic benchmark around a minimal
question: whether a smooth two-term \(\epsilon(p)\) relation can retain the
simplicity of a one-term polytrope while separating canonical-star observables
from high-mass support.  The analysis combines this two-term relation in
\(\epsilon(p)\), an explicit one-term control case, uniform scans with explicit
selection filters, and Tolman--Oppenheimer--Volkoff (TOV)/Love diagnostics.

The remainder of this paper is organized as follows.
Section~\ref{sec:model} defines the double-polytrope EOS and its
thermodynamic consistency conditions.  Section~\ref{sec:stellar-calculations}
summarizes the relativistic stellar-structure and tidal calculations.
Section~\ref{sec:constraints} describes the observational
filters and curve-integral representative-selection score, and Section~\ref{sec:scans} gives
the parameter-space scans and benchmark-selection procedure.
Section~\ref{sec:results} presents the EOS, mass-radius, and tidal results.
Section~\ref{sec:discussion} discusses the scope of the construction
and summarizes the main conclusions.  Additional numerical settings, scan
tables, and Monte Carlo diagnostics are collected in
Appendices~\ref{app:numerical-details}--\ref{app:mc-diagnostics}.

\section{Double-polytrope model}
\label{sec:model}

The model is specified directly as an analytic relation between pressure and
total energy density.  We define the solar compactness length and the
corresponding density scales by
\begin{align}
  r_\odot &= \frac{G\Msun}{c^2}, &
  \rho_\odot &= \frac{\Msun}{r_\odot^3},\\
  \epsilon_\odot &= \rho_\odot c^2, &
  p_\odot &= \epsilon_\odot ,
\end{align}
The dimensionless variables are
\begin{equation}
  \phat = \frac{p}{p_\odot},\qquad
  \ehat = \frac{\epsilon}{\epsilon_\odot}.
\end{equation}
The smooth double-polytrope EOS is then
\begin{equation}
  \ehat(\phat)=a_1\phat^{\Gamma_1}+a_2\phat^{\Gamma_2},
  \label{eq:eos}
\end{equation}
with \(a_i>0\) and \(\Gamma_i>0\).  Unlike a piecewise polytrope, this form has
no internal matching density: both terms contribute at all pressures.  The
parameters are treated phenomenologically and are not mapped directly to
nuclear saturation properties or a microscopic composition.
For stellar-structure calculations, the two monomials provide two effective
control directions over the finite pressure interval sampled by stable stars,
without introducing an internal matching density.  This feature is tested
explicitly against the one-term baseline in Sec.~\ref{sec:results}.

There is no separate crust-core construction.  The same analytic relation is
used down to the numerical pressure surface, and the exterior spacetime is
matched there to Schwarzschild.  This prescription fixes the low-pressure
boundary condition used in all stellar-sequence calculations.  The surface
pressure used in the integrations is many orders of magnitude smaller than the
central pressures of the reported stars, so it has negligible impact on the quoted global quantities at the
precision relevant here.

The thermodynamic derivative is analytic,
\begin{equation}
  \frac{d\ehat}{d\phat}
  =\sum_i a_i\Gamma_i\phat^{\Gamma_i-1},
  \label{eq:dedp}
\end{equation}
and is positive for the parameter domain used in this work.  The model is
therefore barotropically stable in the sense \(dP/d\epsilon>0\).  The
dimensionless sound speed is
\begin{equation}
  \frac{c_s^2}{c^2}
  =\frac{d p}{d\epsilon}
  =\left(\frac{d\ehat}{d\phat}\right)^{-1}.
  \label{eq:sound}
\end{equation}
Causality is enforced over the pressure interval actually sampled by stable
stellar configurations, from the numerical surface through the central pressure
of the maximum-mass model.  The reported stellar observables depend on this
stable-branch pressure domain.

\section{Relativistic stellar calculations}
\label{sec:stellar-calculations}

The stellar model is a static, spherically symmetric perfect fluid with metric
\begin{equation}
  ds^2=-e^{2\Phi(r)}dt^2+
  \left(1-\frac{2m(r)}{r}\right)^{-1}dr^2+r^2d\Omega^2 ,
\end{equation}
where \(G=c=1\) in the differential equations.  Lengths are measured in km in
the numerical integration, and the final mass is converted back to solar masses
using \(G\Msun/c^2\).  For a chosen central pressure \(p_c\), the mass and
pressure profiles obey the Tolman-Oppenheimer-Volkoff equations
\cite{Tolman1939,Oppenheimer1939},
\begin{align}
  \frac{dm}{dr} &= 4\pi r^2\epsilon, \label{eq:tov-m}\\
  \frac{dp}{dr} &=
  -(\epsilon+p)
  \frac{m+4\pi r^3p}{r(r-2m)} . \label{eq:tov-p}
\end{align}
The metric potential is not needed for the mass-radius sequence, but it is
implicitly fixed by
\begin{equation}
  \frac{d\Phi}{dr}=
  \frac{m+4\pi r^3p}{r(r-2m)} ,
\end{equation}
with the exterior Schwarzschild matching condition at the surface.  The
integration starts at a small radius \(r_0\) using the regular central
expansion
\begin{equation}
  m(r_0)=\frac{4\pi}{3}\epsilon_c r_0^3,\qquad p(r_0)=p_c ,
\end{equation}
and terminates when \(p\) reaches the specified surface cutoff.  The
gravitational mass and circumferential radius are \(M=m(R)\) and \(R\).

The stable branch is determined from the one-parameter family of central
pressures.  For a cold barotropic sequence, the Bardeen--Thorne--Meltzer
turning-point criterion identifies the onset of radial instability at the first
maximum of \(M(\rho_c)\) along a fixed-EOS sequence \cite{Bardeen1966}.  We
therefore retain the branch from low central pressure through the first mass
maximum, equivalently the portion with positive \(dM/d\rho_c\) up to the
discrete turning point. 

For each stellar model we also integrate the even-parity quadrupolar tidal
perturbation simultaneously with the TOV equations.  Following the standard
first-order formulation \cite{Hinderer2008,Damour2009}, define \(y(r)\) as the
logarithmic derivative of the radial metric perturbation.  For \(l=2\),
\begin{equation}
  r\frac{dy}{dr}+y^2+yF(r)+r^2\mathcal{U}(r)=0,
  \label{eq:y-tidal}
\end{equation}
with regular central condition \(y(0)=2\).  In the conventions of
Eqs.~\eqref{eq:tov-m} and \eqref{eq:tov-p},
\begin{align}
  F(r) &=
  \frac{1-4\pi r^2(\epsilon-p)}{1-2m/r},\\
  \mathcal{U}(r) &=
  \frac{4\pi}{1-2m/r}
  \left[5\epsilon+9p+\frac{\epsilon+p}{c_s^2}\right]
  -\frac{6}{r^2(1-2m/r)}
  \notag\\
  &\quad
  -4\left[
    \frac{m+4\pi r^3p}{r^2(1-2m/r)}
  \right]^2 .
  \label{eq:u-tidal}
\end{align}
Here \(c_s^2=dp/d\epsilon\) is the dimensionless sound speed in geometrized
units.  The surface value \(y_R=y(R)\), together with the compactness
\(C=M/R\), determines the relativistic quadrupolar Love number.

\begin{align}
  k_2 &=
  \frac{
  8C^5(1-2C)^2\left[2+2C(y_R-1)-y_R\right]
  }{5\mathcal{D}}, \label{eq:k2}\\
  \begin{split}
  \mathcal{D} &=
  2C\left[6-3y_R+3C(5y_R-8)\right]\\
  &\quad
  +4C^3\left[13-11y_R+C(3y_R-2)\right.\\
  &\qquad\left.+2C^2(1+y_R)\right]\\
  &\quad
  +3(1-2C)^2\left[2-y_R+2C(y_R-1)\right]\ln(1-2C).
  \end{split}
\end{align}
The dimensionless tidal deformability reported in the tables and figures is
\begin{equation}
  \Lambda=\frac{2}{3}k_2 C^{-5}.
  \label{eq:lambda}
\end{equation}
Equations~\eqref{eq:y-tidal}--\eqref{eq:lambda} are evaluated only for
configurations satisfying the same stable-branch selection as the mass-radius
sequence.

\section{Observational constraints and curve-integral scoring}
\label{sec:constraints}

Table~\ref{tab:constraints} summarizes the observational and physical filters
that define the benchmark-selection windows.  The mass filter is anchored to
the Shapiro-delay mass of PSR~J0740+6620 \cite{Fonseca2021}.  The
\(\Lambda_{1.4}\) window uses the canonical GW170817 scale, represented here by
the approximate 90\% interval \(190^{+390}_{-120}\), i.e.,
\(70\le\Lambda_{1.4}\le580\), from the Laser Interferometer
Gravitational-Wave Observatory (LIGO)/Virgo common-EOS analysis and the
canonical-deformability summary of Kumar and Landry
\cite{Abbott2018EOS,KumarLandry2019}; related tidal-deformability systematics
are discussed by Zhao and Lattimer \cite{Zhao2018}.  The \(R_{1.4}\) and
\(R_{2.08}\) windows span
current GW/NICER-compatible radii, including the updated PSR~J0740+6620
measurement \(R=12.92^{+2.09}_{-1.13}\) km \cite{Dittmann2024} and the
PSR~J0437--4715 result \(M=1.418\pm0.037\,\Msun\),
\(R=11.36^{+0.95}_{-0.63}\) km \cite{Choudhury2024}.

\begin{table}[htbp!]
\caption{Observational and physical filters used in the final analysis.  The
windows define the selection criteria derived from the cited measurements.}
\label{tab:constraints}
\centering
\footnotesize
\begin{tabular}{@{}ll@{}}
\hline\hline
Constraint & Adopted constraint window  \\
\hline
Maximum mass & $M_{\max} \geq 2.08\,M_\odot$  \\
Canonical radius & $10.5 \leq R_{1.4}/{\rm km} \leq 13.4$  \\
Canonical tidal deformability & $70 \leq \Lambda_{1.4} \leq 580$  \\
High-mass radius & $11.6 \leq R_{2.08}/{\rm km} \leq 15.1$  \\
Causality & $\max_{\rm stable} c_s^2/c^2 \leq 1$ \\
\hline\hline
\end{tabular}
\end{table}

Within the viable regions defined by these filters, representative EOSs are
ranked with a curve-integral score constructed from public observational
posterior data sets.  For a parameter vector \(\bm\theta\), the
TOV/Love solver gives a stable branch \(R_\theta(M)\) and
\(\Lambda_\theta(M)\), together with \(M_{\max,\theta}\) and the maximum
stable-branch sound speed.  The score used for representative selection is
\begin{equation}
  S(\bm\theta)
  =
  S_{\rm NICER}(\bm\theta)
  +S_{\rm GW170817}(\bm\theta)
  +S_{\rm mass}(\bm\theta)
  +S_{\rm phys}(\bm\theta).
  \label{eq:empirical-curve-score}
\end{equation}
The NICER term integrates the model mass-radius curve through the public
PSR~J0437--4715 and PSR~J0740+6620 posterior samples~\citep{Choudhury2024,Dittmann2024},
\begin{align}
  S_{\rm NICER}
  &=
  \sum_i
  \log
  \left[
  \frac{1}{\Delta M_i}
  \int_{\mathcal M_i}
  p_i\!\left(M,R_\theta(M)\right)\,dM
  \right]
  \notag\\
  &\quad+\sum_i\log f_i .
  \label{eq:nicer-curve-score}
\end{align}
where \(p_i(M,R)\) is a smoothed two-dimensional density from those
public samples, \(\mathcal M_i\) is the overlap between the model's stable mass
range and the support of source \(i\), and \(f_i\) is the corresponding overlap
fraction.  The GW170817 term uses the public parametrized-EOS posterior
samples~\citep{Abbott2018EOS} to score the model's
\(\Lambda_\theta(M)\) curve:
\begin{align}
  S_{\rm GW170817}
  &=
  \log
  \left\langle
  p_1\!\left[\log\Lambda_\theta(m_1)\mid m_1\right]\right.
  \notag\\
  &\qquad\left.
  \times
  p_2\!\left[\log\Lambda_\theta(m_2)\mid m_2\right]
  \right\rangle_{(m_1,m_2)}
  +\log f_{\rm GW}.
  \label{eq:gw-curve-score}
\end{align}
Here the average is over the public GW170817 mass pairs whose component masses
are covered by the model sequence, and the conditional densities are obtained
from smoothed distributions in \((m,\log\Lambda)\).  The massive-pulsar
term is
\begin{equation}
  S_{\rm mass}
  =
  \sum_j
  \log\Phi
  \left[
  \frac{M_{\max,\theta}-M_j}{\sigma_j}
  \right],
  \label{eq:mass-score}
\end{equation}
with the three mass measurements used in Table~\ref{tab:constraints}.  Finally,
\(S_{\rm phys}\) assigns a finite penalty to numerical failures and to stable
branches approaching the causal limit; models with
\(\max_{\rm stable}c_s^2/c^2>1\) are excluded from the representative candidate set.

Equations~\eqref{eq:empirical-curve-score}--\eqref{eq:mass-score} rank analytic
EOSs by their overlap with the mass-radius and mass-deformability regions shown
in Figs.~\ref{fig:mass-radius} and~\ref{fig:lambda-mass}.

\section{Parameter-space scans and representative selection}
\label{sec:scans}

The double-polytrope parameters are
\begin{equation}
  \bm\theta=(\Gamma_1,\Gamma_2,\log_{10}a_1,\log_{10}a_2).
\end{equation}
Sobol low-discrepancy scans are used to map filter feasibility across the
chosen parameter ranges.  A one-term baseline,
\begin{equation}
  \ehat(\phat)=a\phat^\Gamma ,
\end{equation}
is scanned first to test whether the simplest analytic ansatz is already
sufficient.  The two-term model is then scanned in the chosen parameter
domain.  Because the two terms in
Eq.~\eqref{eq:eos} can be interchanged without changing the EOS, the full
four-dimensional parameter space contains two equivalent feasible regions.  We
report one ordered representative of this exchange-symmetric pair by imposing
\(\Gamma_1<\Gamma_2\).  The lower bound \(\Gamma_1>0.1\) is also imposed so
that the first term does not degenerate into an approximately constant
contribution over the stellar pressure interval.  All scan points are evaluated
with the same TOV/Love solver and classified against
Table~\ref{tab:constraints}.  The explicit scan boxes, sample counts, and
numerical settings are
given in Appendix~\ref{app:numerical-details}.

To select representative curves within the feasible region, a Monte Carlo
calculation samples the same analytic parameterization using adaptive
Metropolis updates and differential-evolution proposals
\cite{Haario2001,TerBraak2006}.  The candidates are then rescored with the
curve-integral score in Eq.~\eqref{eq:empirical-curve-score};
the final run used a random subset of 20,000 candidates from a 320,000-sample
pool.  These rescored candidates define a score-weighted distribution over the
analytic ansatz and identify the maximum-score sample and the componentwise
score-weighted mean and median models, which are then re-evaluated at higher
TOV/Love resolution.

\section{Results}
\label{sec:results}

The one-term baseline scan finds no accepted model.  The rejected
models show a consistent pattern across several diagnostics: points that satisfy the adopted
\(R_{1.4}\) and \(\Lambda_{1.4}\) windows can also approach the high-mass
radius window, but the subset that also satisfies the \(M_{\max}\) and
\(R_{2.08}\) filters is acausal on the stable branch.
This behavior reflects the limited control afforded by
\(\ehat=a\phat^\Gamma\): the same coefficient and exponent determine both the
intermediate-pressure stiffness that fixes canonical radii and tides and the
higher-pressure stiffness that supports the most massive stable configurations.
In the subset highlighted in Fig.~\ref{fig:singlepoly-scan}, 59 points satisfy the canonical
\(R_{1.4}\)-\(\Lambda_{1.4}\) windows and 23 of these also satisfy the
adopted \(M_{\max}\) and \(R_{2.08}\) filters, but their stable-branch maxima
lie in the range \(\max c_s^2/c^2=2.23\)--3.76.  Figure~\ref{fig:singlepoly-scan}
therefore shows that the one-term ansatz is close in some projections but does
not provide an acceptable causal analytic EOS in the chosen scan box.  The
double-polytrope ansatz is the minimal extension considered in the remainder of
the analysis.

\wideinlinefigure{fig:singlepoly-scan}{0.7\textwidth}{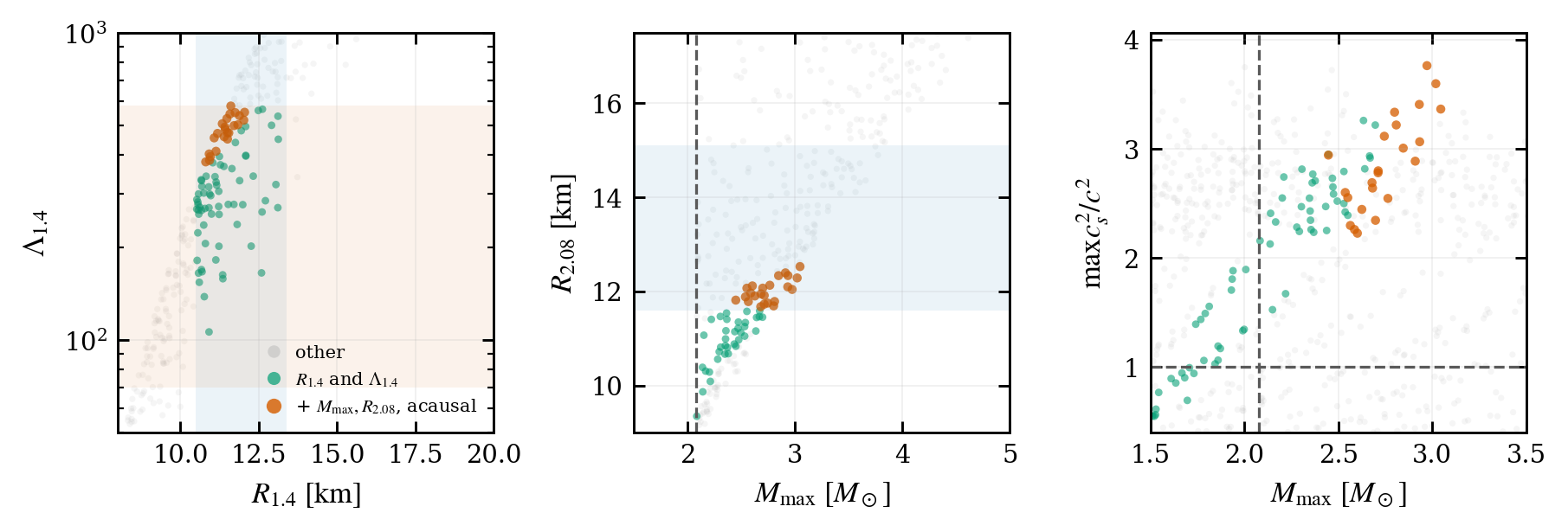}
{One-term baseline feasibility scan in observable space.  Panel (a)
shows the \(R_{1.4}\)-\(\Lambda_{1.4}\) projection; green and orange points
enter the canonical radius--tidal window.  Panel (b) projects the same
canonical-window candidates into the \(M_{\max}\)-\(R_{2.08}\) plane; orange
points also satisfy the adopted high-mass filters.  Panel (c) shows the
corresponding \(M_{\max}\)-\(\max c_s^2/c^2\) projection.  All orange points lie
above the causal bound \(c_s^2/c^2=1\), so no one-term model passes all filters in the scanned parameter box.}

The double-polytrope scan locates a viable portion of the
four-dimensional ordered box.  The expanded 300,000-point Sobol scan described
in Appendix~\ref{app:numerical-details} yields 318 accepted models, forming the
parameter-space and observable-space structures shown in
Figs.~\ref{fig:double-scan} and \ref{fig:double-scan-observables}.

\wideinlinefigure{fig:double-scan}{0.8\textwidth}{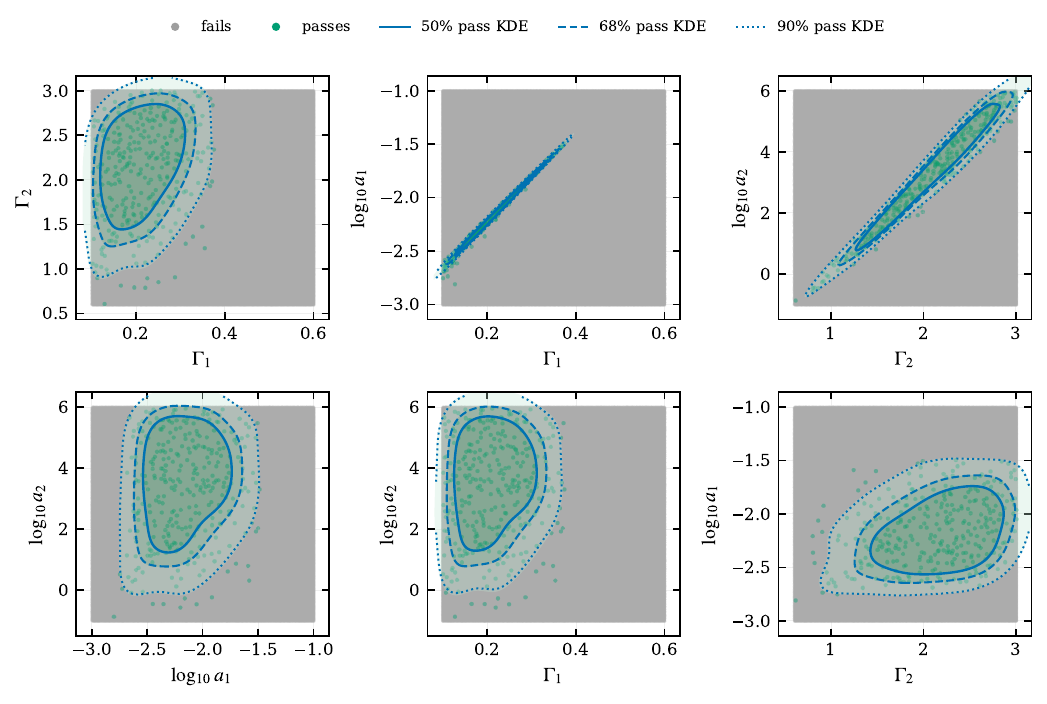}
{Double-polytrope feasibility scan in parameter space.
Grey points are numerically converged filter rejections and green points pass all
filters.  Blue curves mark
50\%, 68\%, and 90\% Gaussian KDE highest-density regions of the passing
points.  The second exchange-symmetric solution region is obtained by
interchanging the two polytropic terms.}

\wideinlinefigure{fig:double-scan-observables}{0.7\textwidth}{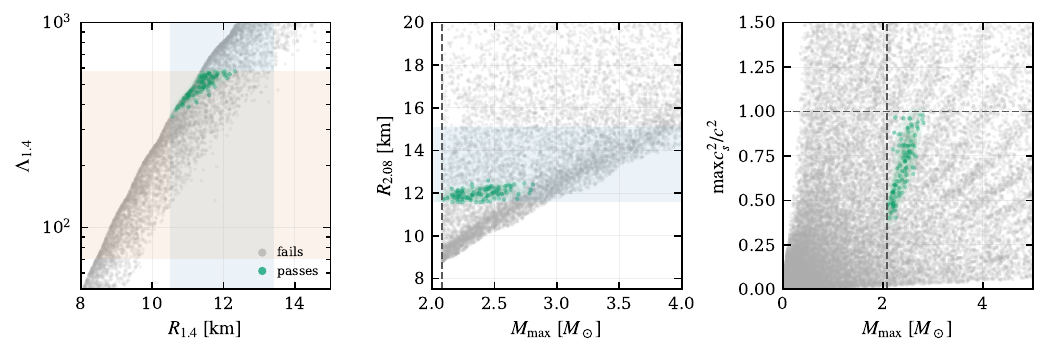}
{Observable-space projection of the double-polytrope scan.
	Colors identify the filter status of each numerically converged model.
	Green points mark models that satisfy the adopted \(R_{1.4}\),
	\(\Lambda_{1.4}\), \(M_{\max}\), \(R_{2.08}\), and stable-branch causality
	requirements simultaneously.}

The nearly linear ridges in Fig.~\ref{fig:double-scan} trace parameter
degeneracy directions imposed by the finite pressure range sampled by stable stars.  In
logarithmic form, each EOS contribution satisfies
\(\log\epsilon_i=\log a_i+\Gamma_i\log p\).  If a given mass range primarily
probes pressures near an effective pivot \(p_\ast\), keeping the corresponding
energy-density contribution approximately fixed requires
\(\Delta\log a_i\simeq-\log p_\ast\,\Delta\Gamma_i\).  The accepted scan cloud
therefore follows approximately straight tracks in the
\((\Gamma_i,\log_{10}a_i)\) projections.  The lower-exponent pair
\((\Gamma_1,a_1)\) is mainly constrained by the pressure interval that controls
\(R_{1.4}\) and \(\Lambda_{1.4}\), whereas \((\Gamma_2,a_2)\) is constrained by
the higher-pressure interval needed to support the \(2.08\,\Msun\) scale while
remaining causal.  Figure~\ref{fig:double-scan-observables} shows the
observable consequence of the same degeneracy: many parameter combinations lie
near the canonical-radius and tidal windows, but only a subset also has enough
high-density stiffness without exceeding \(c_s^2/c^2=1\).  Related
parameterization-induced correlations have been discussed in broader EOS inference settings
\cite{Legred2022}.

The accepted double-polytrope scan points have
\(R_{1.4}=10.99\)--\(11.73\) km and
\(\Lambda_{1.4}=432\)--556 over the central 68\% of the accepted scan cloud.
The corresponding high-mass radii are
\(R_{2.08}=11.76\)--\(12.33\) km, and the maximum stable-branch sound speeds
span \(0.54\)--\(0.91\).  These intervals summarize the scale of the accepted
cloud.

The representative-selection Monte Carlo sample defines three benchmark EOSs:
the maximum-score representative, labeled MAP in analogy with the conventional
maximum-a-posteriori abbreviation, the componentwise score-weighted median, and
the componentwise score-weighted mean.  Their parameters are
shown in Table~\ref{tab:rep-params}.  All three pass the filters after
high-resolution re-evaluation and serve as reference points within the
mapped analytic feasible region.

\begin{table}[t]
	\caption{Representative double-polytrope parameters in
		Eq.~\eqref{eq:eos}.}
	\label{tab:rep-params}
	\centering
	\footnotesize
	\begin{tabular}{lcccc}
		\hline\hline
		Model & $\Gamma_1$ & $\Gamma_2$ & $a_1$ & $a_2$ \\
		\hline
		MAP & 0.19137 & 1.61011 & 0.00533 & 84.47086 \\
		median & 0.22713 & 2.20319 & 0.00783 & 3743.13781 \\
		mean & 0.22835 & 2.09076 & 0.00786 & 1958.45576 \\
		\hline\hline
	\end{tabular}
\end{table}

\begin{table}[t]
\caption{High-resolution TOV/Love diagnostics for representative models.}
\label{tab:rep-observables}
\centering
\footnotesize
\begin{tabular}{lccccc}
\hline\hline
Model & $M_{\max}$ & $R_{1.4}$ & $\Lambda_{1.4}$ & $R_{2.08}$ & $\max c_s^2/c^2$ \\
 & $[M_\odot]$ & [km] & & [km] & \\
\hline
MAP & 2.444 & 11.329 & 512.1 & 12.165 & 0.594 \\
median & 2.491 & 11.317 & 484.5 & 12.102 & 0.711 \\
mean & 2.446 & 11.357 & 494.0 & 12.115 & 0.656 \\
\hline\hline
\end{tabular}
\end{table}

The maximum-score representative
\begin{equation}
  \ehat(\phat)=0.00532781\,\phat^{0.191371}
  +84.4709\,\phat^{1.61011},
\end{equation}
gives \(M_{\max}=2.444\,\Msun\), \(R_{1.4}=11.329\) km,
\(\Lambda_{1.4}=512.1\), and \(R_{2.08}=12.165\) km.  Its maximum stable-branch
sound speed is \(0.594\,c^2\).  The score-weighted median and mean preserve
nearby canonical-radius and tidal-deformability values while sampling different
high-pressure compensation directions in the analytic ansatz.
Additional stable-branch diagnostics, including
canonical compactness and central-density values, are listed in
Appendix~\ref{app:numerical-details}.

Figure~\ref{fig:eos-pressure} shows the pressure-energy relation for the three
representative EOSs over the pressure interval
\(10^{-6}\le p/p_\odot\le2\times10^{-3}\).  The grey curves show the SLy and
H4 Read et al.\ piecewise-polytrope fits as reference scales.  The
representative curves lie on comparable pressure scales over the displayed
density interval, with visible separation at high density.

\begin{figure}[t]
	\centering
	\includegraphics[width=0.9\columnwidth]{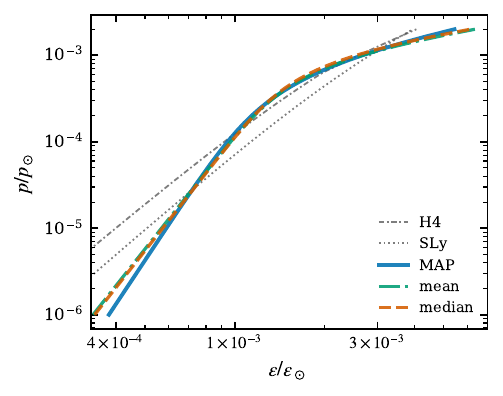}
	\caption{Representative EOS pressure as a function of total energy density
		for the maximum-score, score-weighted median, and score-weighted mean
		models.  The grey curves show the SLy and H4 Read et al.\
		piecewise-polytrope fits.}
	\label{fig:eos-pressure}
\end{figure}

\begin{figure}[t]
	\centering
	\includegraphics[width=0.9\columnwidth]{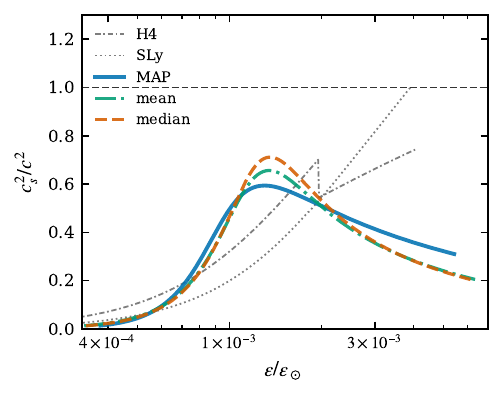}
	\caption{Representative stable-branch sound speeds with the SLy and H4 Read
		et al.\ piecewise-polytrope fits shown in grey.  The dashed line marks
		\(c_s^2/c^2=1\).}
	\label{fig:sound-speed}
\end{figure}

The corresponding sound-speed profiles are shown in
Fig.~\ref{fig:sound-speed}.  The maximum-score model has
\(\max c_s^2/c^2=0.594\), while the score-weighted median and mean have
\(\max c_s^2/c^2=0.711\) and 0.656.  In all three cases the stable branch remains
well below the causal bound.  The sharp features in the reference curves, most
visibly in H4, reflect derivative discontinuities of the piecewise-polytrope
fits and provide comparison scales.

\begin{figure}[t]
	\centering
	\includegraphics[width=0.9\columnwidth]{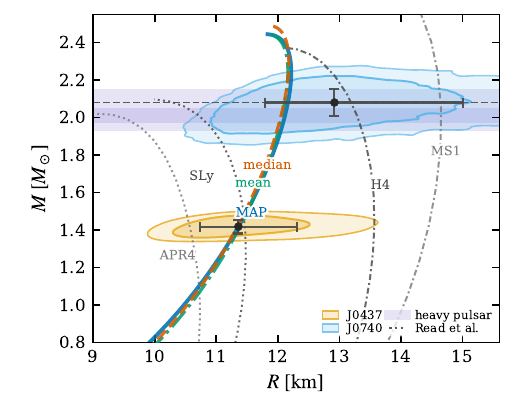}
	\caption{Representative stable mass-radius sequences with displayed
		Neutron Star Interior Composition Explorer (NICER) mass-radius
		highest-posterior-density regions: the orange/yellow
		region shows PSR~J0437--4715 and the blue region shows PSR~J0740+6620.  Purple
		horizontal bands mark heavy-pulsar masses.  The black marker gives the
		published PSR~J0437--4715 headline mass-radius interval.  Thin grey
		patterned curves, labeled by EOS name, are selected Read et al.\
		piecewise-polytrope fits shown as references.}
	\label{fig:mass-radius}
\end{figure}

\begin{figure}[t]
	\centering
	\includegraphics[width=0.9\columnwidth]{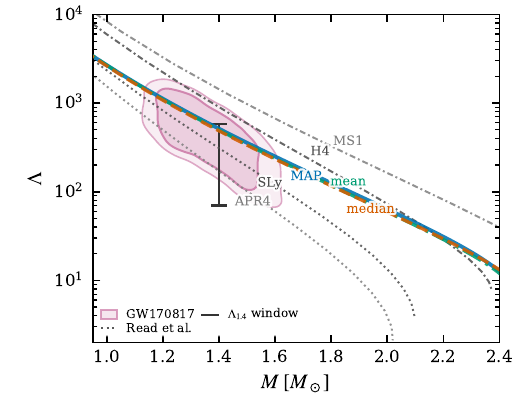}
	\caption{Representative dimensionless tidal deformability along the stable
		branch.  The shaded region is the GW170817 component \(M\)-\(\Lambda\)
		highest-posterior-density projection from the public parametrized-EOS
		posterior samples.  The vertical bracket at \(1.4\,\Msun\) marks the
		adopted \(\Lambda_{1.4}\) screening window; thin grey patterned curves are
		the same Read et al.\ reference fits as in Fig.~\ref{fig:mass-radius}.}
	\label{fig:lambda-mass}
\end{figure}

Figure~\ref{fig:mass-radius} displays the stable mass-radius sequences.  The
thin grey reference curves are the SLy, APR4, H4, and MS1 four-parameter
piecewise-polytrope fits of Read et al.~\cite{Read2009}.  They are included as
familiar realistic-EOS scale references: SLy and APR4 provide relatively
compact examples, whereas H4 and MS1 illustrate stiffer behavior.  The orange
region shows the public PSR~J0437--4715 NICER 68\% and 95\% mass-radius
posterior contours of Choudhury et al.~\cite{Choudhury2024}.  The blue region
is computed from the public PSR~J0740+6620 NICER/XMM-Newton mass-radius posterior
samples of Dittmann et al.~\cite{Dittmann2024}; the two nested contours
enclose the smoothed 68\% and 90\% highest-posterior-density regions.  The
black PSR~J0437--4715 marker and the PSR~J1614--2230, PSR~J0348+0432, and
PSR~J0740+6620 high-mass bands are retained as visual guides.  The
representative curves pass through both the low-mass and high-mass NICER
regions while sharing a narrow canonical-radius band near
\(R_{1.4}\simeq 11.3\) km.  Their maximum masses span
\(2.444\)--\(2.491\,\Msun\).

The corresponding mass-deformability sequences are shown in
Fig.~\ref{fig:lambda-mass}.  The pink region is obtained from the public
GW170817 parametrized-EOS posterior samples \cite{Abbott2018EOS}.  The representative
\(\Lambda(M)\) curves cross this region and also satisfy the displayed
\(\Lambda_{1.4}\) screening bracket, with
\(\Lambda_{1.4}=485\)--512.  The reference fits span both softer and stiffer
mass--tidal behavior and provide comparison scales for the analytic
benchmarks.
Appendix~\ref{app:mc-diagnostics} gives the representative-selection Monte
Carlo diagnostics.

\section{Discussion and summary}
\label{sec:discussion}

The scans show that the one-term baseline is too restrictive in the adopted
parameter box, while the smooth two-term relation provides accepted causal
models.  The representatives reported in
Tables~\ref{tab:rep-params} and
\ref{tab:rep-observables} have \(M_{\max}=2.44\)--\(2.49\,M_\odot\),
\(R_{1.4}\simeq11.3\) km, \(\Lambda_{1.4}=485\)--512, and stable-branch sound
speeds below the causal limit.

The benchmark can serve as a reproducible test EOS for TOV and Love-number
solvers, convergence and interpolation studies, validation of semi-analytic
approximations, and sensitivity tests in which a tabulated microphysical EOS
would make the EOS-dependence harder to isolate.  The analytic form also gives
direct access to \(d\epsilon/dp\) and the sound speed, making thermodynamic
stability and stable-branch causality checks straightforward.

The filters and curve-integral score used here should not be read as a full
joint posterior EOS inference.  They define a reproducible
benchmark-selection procedure over the chosen analytic parameterization.  The
resulting EOS is therefore aimed at relativistic stellar-structure tests rather
than crust physics, low-mass neutron-star modeling, thermal evolution,
composition inference, or the extraction of nuclear-matter parameters.  Those
applications require microphysical information that is deliberately not encoded
in the two-term analytic \(\epsilon(p)\) relation.  Within this intended role,
the model supplies an observationally consistent analytic reference case whose
assumptions are explicit, whose stellar sequences are reproducible from the
archived scripts, and whose cost is low enough for repeated use in exploratory
calculations.

\vspace{0.5\baselineskip}
\noindent\textbf{Acknowledgments.}
The authors thank their colleges for everyday discussions. K.Z. (Hong Zhang) is supported by a classified fund from Shanghai city.

\bibliographystyle{aasjournal}
\bibliography{CZZ}

\appendix
\setcounter{figure}{0}
\setcounter{table}{0}
\renewcommand{\thefigure}{A\arabic{figure}}
\renewcommand{\thetable}{A\arabic{table}}
\renewcommand{\theHfigure}{appendix.\arabic{figure}}
\renewcommand{\theHtable}{appendix.\arabic{table}}

\section{Numerical settings}
\label{app:numerical-details}

The one-term baseline uses a uniform Sobol scan in
\(\Gamma\in[0.20,1.40]\) and \(\log_{10}a\in[-3.00,2.00]\).  The
double-polytrope scan used for the final feasible-region figures is uniform in
the ordered box
\[
\begin{aligned}
  \Gamma_1 &\in [0.10, 0.60],\\
  \Gamma_2 &\in [0.60, 3.00],\\
  \log_{10}a_1 &\in [-3.00, -1.00],\\
  \log_{10}a_2 &\in [-1.00, 6.00].
\end{aligned}
\]
The ordering convention selects the \(\Gamma_1<\Gamma_2\) representative of the
exchange-symmetric parameter space, while the \(\Gamma_1>0.1\) lower cutoff
prevents the low-exponent term from acting as a nearly pressure-independent
offset over the relevant stellar-pressure range.
The final Sobol scan contains 300,000 points and 318 accepted models.
The representative EOSs are re-evaluated with 220 central-pressure points, a
maximum radial step of \(0.01\) km, and 1600 sound-speed samples over the
stable pressure domain.  The EOS and sound-speed panels are drawn over
\(10^{-6}\le p/p_\odot\le2\times10^{-3}\), which covers the relevant stable
stellar configurations for the accepted models and avoids displaying
high-pressure extrapolation outside the stable branch.  The transparent interval in
the auxiliary EOS--sound-speed diagnostic uses all 318 accepted
Sobol points for the EOS envelope; mass-sequence envelopes use 96 selected accepted
models evaluated on a common mass grid.

\section{Auxiliary scan and representative-selection tables}
\label{app:aux-tables}

\renewcommand{\arraystretch}{0.92}

\begin{center}
\refstepcounter{table}\label{tab:scan-summary}
\begin{minipage}{0.98\columnwidth}
\footnotesize
\textbf{Table~\thetable.} Summary of the blind parameter-space scans.
Dominant failure labels are nonexclusive and refer to the largest fractions
among rejected points: numerical/interpolation, high \(\Lambda_{1.4}\), and
causality for the single-polytrope baseline; numerical/interpolation, high
\(\Lambda_{1.4}\), and high \(R_{1.4}\) for the double-polytrope
scan.
\end{minipage}

\vspace{3pt}
\footnotesize
\begin{tabular}{@{}lrrr@{}}
\hline\hline
Scan & Samples & Passed & Pass fraction \\
\hline
Single-polytrope baseline & 20,000 & 0 & 0\% \\
Double-polytrope parameter box & 300,000 & 318 & 0.106\% \\
\hline\hline
\end{tabular}
\end{center}

\begin{center}
\refstepcounter{table}\label{tab:scan-pass-quantiles}
\begin{minipage}{0.98\columnwidth}
\footnotesize
\textbf{Table~\thetable.} Central 68\% intervals of the accepted
double-polytrope scan points.  These intervals summarize the
feasibility scan.
\end{minipage}

\vspace{3pt}
\footnotesize
\begin{tabular}{lccc}
\hline\hline
Quantity & 16\% & Median & 84\% \\
\hline
$\Gamma_1$ & 0.13625 & 0.21049 & 0.29170 \\
$\Gamma_2$ & 1.5143 & 2.1781 & 2.6782 \\
$\log_{10}a_1$ & -2.4864 & -2.1916 & -1.8354 \\
$\log_{10}a_2$ & 1.4565 & 3.5265 & 5.1884 \\
$R_{1.4}$ & 10.993 & 11.367 & 11.728 \\
$\Lambda_{1.4}$ & 431.52 & 497.16 & 556.30 \\
$R_{2.08}$ & 11.760 & 12.006 & 12.327 \\
$\max c_s^2/c^2$ & 0.54137 & 0.72273 & 0.90926 \\
\hline\hline
\end{tabular}
\end{center}

\begin{table*}[b]
\caption{Representative-selection parameter estimates under the
curve-integral score in Eq.~\eqref{eq:empirical-curve-score}.
Intervals are central 68\% intervals of the retained score-weighted sample.}
\label{tab:posterior-params}
\centering
\footnotesize
\begin{tabular}{lccc}
\hline\hline
Parameter & Mean & Median & 68\% interval \\
\hline
$\Gamma_1$ & 0.22835 & 0.22713 & $[0.15404, 0.30295]$ \\
$\Gamma_2$ & 2.09076 & 2.20319 & $[1.47315, 2.65760]$ \\
$a_1$ & 0.00983 & 0.00783 & $[0.00373, 0.01627]$ \\
$a_2$ & \(6.49694\times10^4\) & 3743.13808 & $[23.10326, 9.93378\times10^4]$ \\
\hline\hline
\end{tabular}
\end{table*}

\begin{table}[t]
\caption{Diagnostics for the representative-selection curve-rescoring run.}
\label{tab:mcmc-diag}
\centering
\footnotesize
\begin{tabular}{lc}
\hline\hline
Diagnostic & Value \\
\hline
Candidate pool samples & 320000 \\
Curve-rescored samples & 20000 \\
Valid curve evaluations & 20000 \\
Importance effective sample size (ESS) & 392.8 \\
\hline\hline
\end{tabular}
\end{table}

\begin{table*}[t]
\caption{Additional stable-branch physical diagnostics for the representative
models.  The central mass densities are quoted at \(1.4\,\Msun\) and at the
maximum-mass central configuration.}
\label{tab:physical}
\centering
\scriptsize
\begin{tabular}{lccccc}
\hline\hline
Model & $C_{1.4}$ & $k_2(1.4)$ & $\rho_{c,1.4}$ & $\rho_{c,\max}$ & $1-\max c_s^2/c^2$ \\
 & & & [$10^{15}{\rm g\,cm^{-3}}$] & [$10^{15}{\rm g\,cm^{-3}}$] & \\
\hline
MAP & 0.182 & 0.1554 & 0.645 & 1.724 & 0.406 \\
median & 0.183 & 0.1479 & 0.670 & 1.743 & 0.289 \\
mean & 0.182 & 0.1481 & 0.668 & 1.734 & 0.344 \\
\hline\hline
\end{tabular}
\end{table*}

\section{Monte Carlo diagnostics}
\label{app:mc-diagnostics}

\begin{figure*}[t]
	\centering
	\begin{minipage}[c]{0.55\textwidth}
		\centering
		\includegraphics[width=\linewidth]{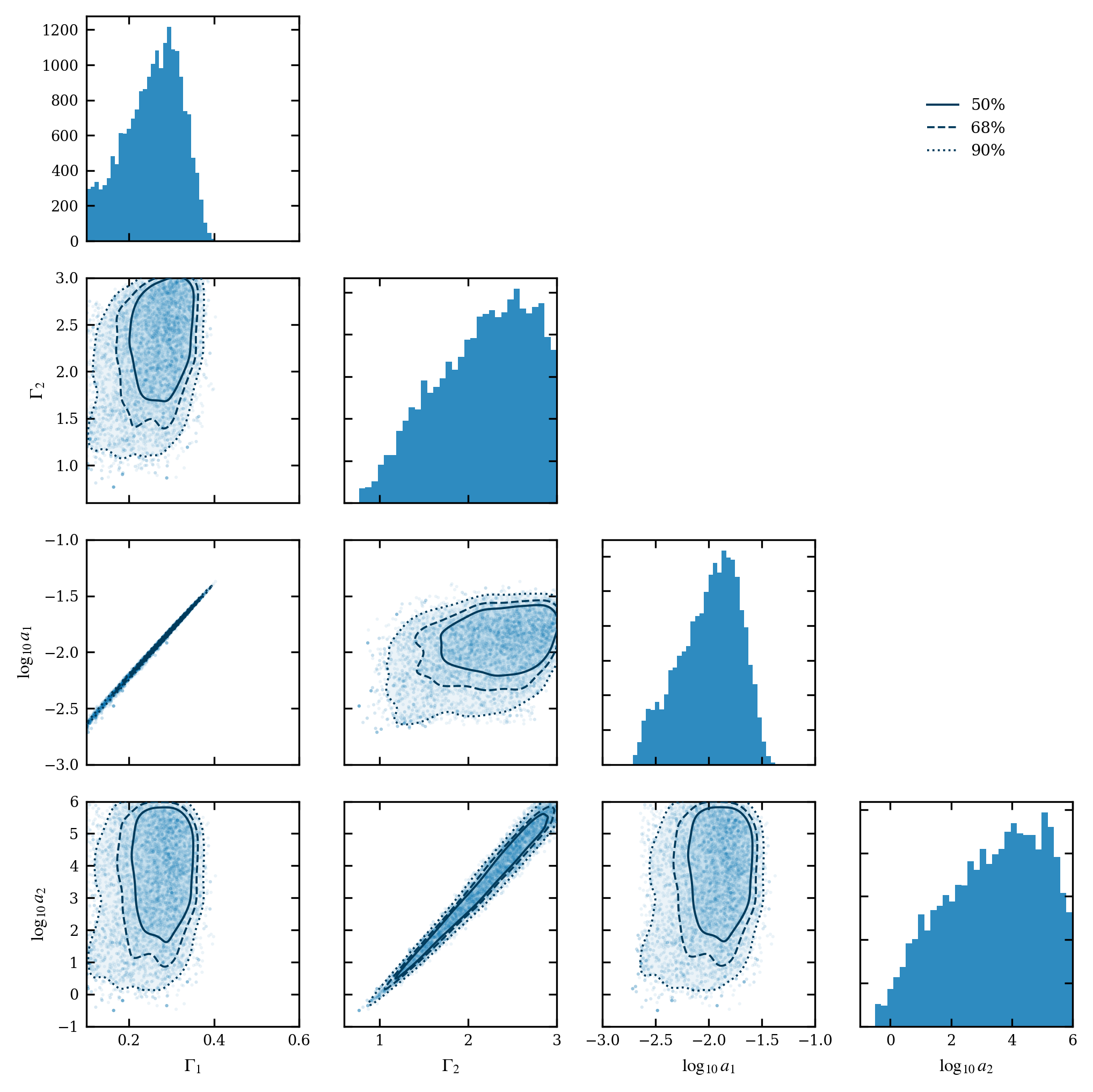}
		\centerline{(a)}
	\end{minipage}\hfill
	\begin{minipage}[c]{0.35\textwidth}
		\centering
		\includegraphics[width=\linewidth]{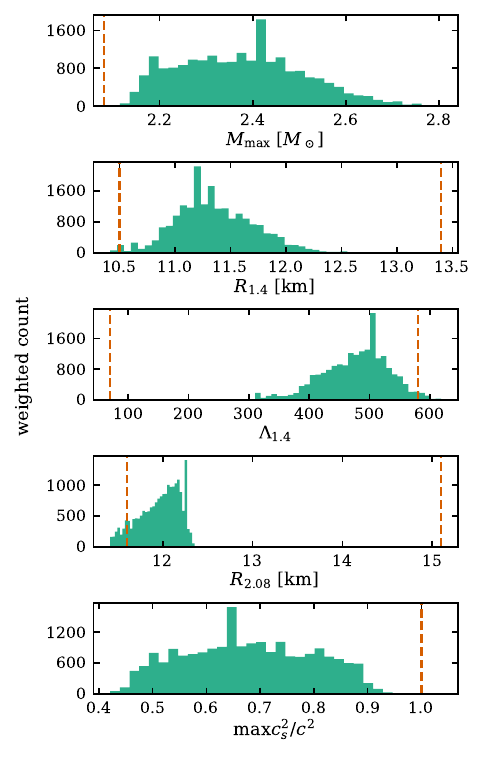}
		\centerline{(b)}
	\end{minipage}
\caption{Representative-selection Monte Carlo diagnostics under the
		curve-integral score in Eq.~\eqref{eq:empirical-curve-score}.  Panel (a) shows parameter samples and
		50\%, 68\%, and 90\% Gaussian KDE highest-density sample-mass contours in
		the same \((\Gamma_1,\Gamma_2,\log_{10}a_1,\log_{10}a_2)\) parameter box as
		Fig.~\ref{fig:double-scan}; panel (b) shows score-weighted one-dimensional
		distributions of \(M_{\max}\), \(R_{1.4}\), \(\Lambda_{1.4}\),
		\(R_{2.08}\), and \(\max c_s^2/c^2\), with orange dashed lines marking the
		relevant filter boundaries.}
	\label{fig:mc-appendix}
\end{figure*}

Figure~\ref{fig:mc-appendix} shows parameter-space diagnostics and the
corresponding observable distributions for the Monte Carlo sample used to
select representative models.  The contours in the two-dimensional panels
enclose fixed fractions of sample mass, estimated as
Gaussian KDE highest-density regions.  They represent
score-weighted probability-mass contours within the adopted analytic
parameterization and scoring function.  The observable distributions summarize
where the same score-weighted sample lies relative to the maximum-mass,
canonical-radius, canonical-deformability, high-mass-radius, and causality
filters.

\end{document}